\documentclass[12pt,aps,preprint,pra]{revtex4}
\usepackage{graphicx}

\begin{document}
\title{Electronic Properties of Boron and Nitrogen doped graphene: A first principles study}
\author{Sugata Mukherjee}
\email{sugata@bose.res.in}
\author{T. P. Kaloni}
\email{tkaloni@gmail.com}
\affiliation{S.N. Bose National Centre for Basic Sciences, Block-JD, Sector III, Salt Lake Kolkata 700097 (India) }

\begin{abstract}
Effect of doping of graphene either by Boron (B), Nitrogen (N) or co-doped by B and N is studied using density 
functional theory. Our extensive band structure and density of states calculations indicate that upon 
doping by N (electron doping), the Dirac point in the graphene band structure shifts below the Fermi level 
and an energy gap appears at the high symmetric K-point. On the other hand, by B (hole doping), the Dirac point shifts 
above the Fermi level and a gap appears. Upon co-doping of graphene by B and N, the energy gap between valence and conduction 
bands appears at Fermi level and the system behaves as narrow gap semiconductor. Obtained results are found to be in well agreement 
with available experimental findings.
\end{abstract}

\date{\today}
\maketitle
\section{Introduction}
\label{intro}
Graphene has been a topic of intensive investigation since the pioneering work on synthesis of 
this unique two-dimensional material by Geim, Novoselov and co-workers \cite{Ref1,Ref2,Ref3}. Recently, various 
attempts have been made to fabricate graphene devices by engineering their band gaps by doping \cite{Ref4,Ref5,Ref6}.  
Investigations on doped graphene nano-ribbons \cite{Ref4,Ref5} indicate that upon doping by N or B 
n-type or p-type semiconducting graphene  can possibly be obtained. It has been experimentally established that 
upon N doping of graphene \cite{Ref4}, the Dirac point in the band structure of graphene tends to move 
below the $E_F$ and an energy gap appears at high-symmetric K-point.
	
Very recently, Ci et al. \cite{Ref6} have been able to synthesize a novel two-dimensional nanomaterial where a 
few Carbon  atoms on a graphene sheet is replaced by equal number of B and N atoms. 
The concentration of the dopant atoms can be controlled by keeping B/N ratio same. This novel BNC nanomaterial 
was found be semiconducting with a very small gap between the valence and conduction bands. Synthesis
of similar BNC materials have also been reported by Panchakarla et al. \cite{Ref7}. The electronic properties of N and B 
doped graphene nanoribbons with armchair edges also has been reported \cite{Ref8}. 
N introduces an impurity level above the donor level, while an impurity level introduced by B is below that of the 
acceptor level. In contrast to single wall carbon nanotubes, the impurity level is neither donor nor acceptor in
their systems. The donor and acceptor levels are derived mainly from the lowest unoccupied orbital and the highest 
occupied orbital.  

Theoretically, there are various possibilities to introduce gap in graphene, i.e. by oxidation of monovacancies in graphene 
\cite{Ref9}, graphene/boron nitride heterobilayers \cite{Ref10,kaloni1}, F-intercalated graphene on SiC 
substrate \cite{Ref11}, and bilayer graphene-BN heterostructures. Experimentally, substitutional carbon doping of boron nitride 
nanosheets, nanoribbons, and nanotubes has been reported \cite{Ref12}. Experimentally, it has been observed that the $sp^2$ 
hybridized BNC nano-structure, with equal number of B and N atoms, can open finite band gap \cite{Ref6}.

Motivated by these current experimental and theoretical reports, we investigated in this paper
the effect of doping of Boron and Nitrogen and also of their co-doping on the electronic properties of the 
graphene systems using first-principles electronic structure calculations based on density functional theory.
Our results indicate that due to the presence of  B or N atoms Graphene is more stable as compared to pristine $h$-BN. 
This suggests that the $h$-BNC domains are more likely to form in $h$-CBN nanomaterial. The electronic properties of Graphene 
with single substitutional B or N atoms, indicative of a $p$ ($n$)-doping, can be tuned easily as reported recently
\cite{prb}. Furthermore, we have carried out extensive band structure calculations \cite{Ref8} 
in the framework of the local spin-density formalism (LSDA) using first-principles 
pseudopotentials \cite{Ref13} and examined the effect of B and N doped multilayers graphene. To best of our knowledge, the stability 
and electronic properties of such a system has not been well understood. We believe our results explain the formation 
of $h$-BNC nano-materials as it has been proven experimentally \cite{Ref6}. These new form of hybridized $h$-BNC nanomaterial enables 
for the development of band gap engineering applications, in particular, in nano-electronics and nano-optics that are distinct 
from those of pristine graphene and pristine $h$-BN.

\section{Calculation Method}
\begin{figure}
\includegraphics[width=1.2\textwidth]{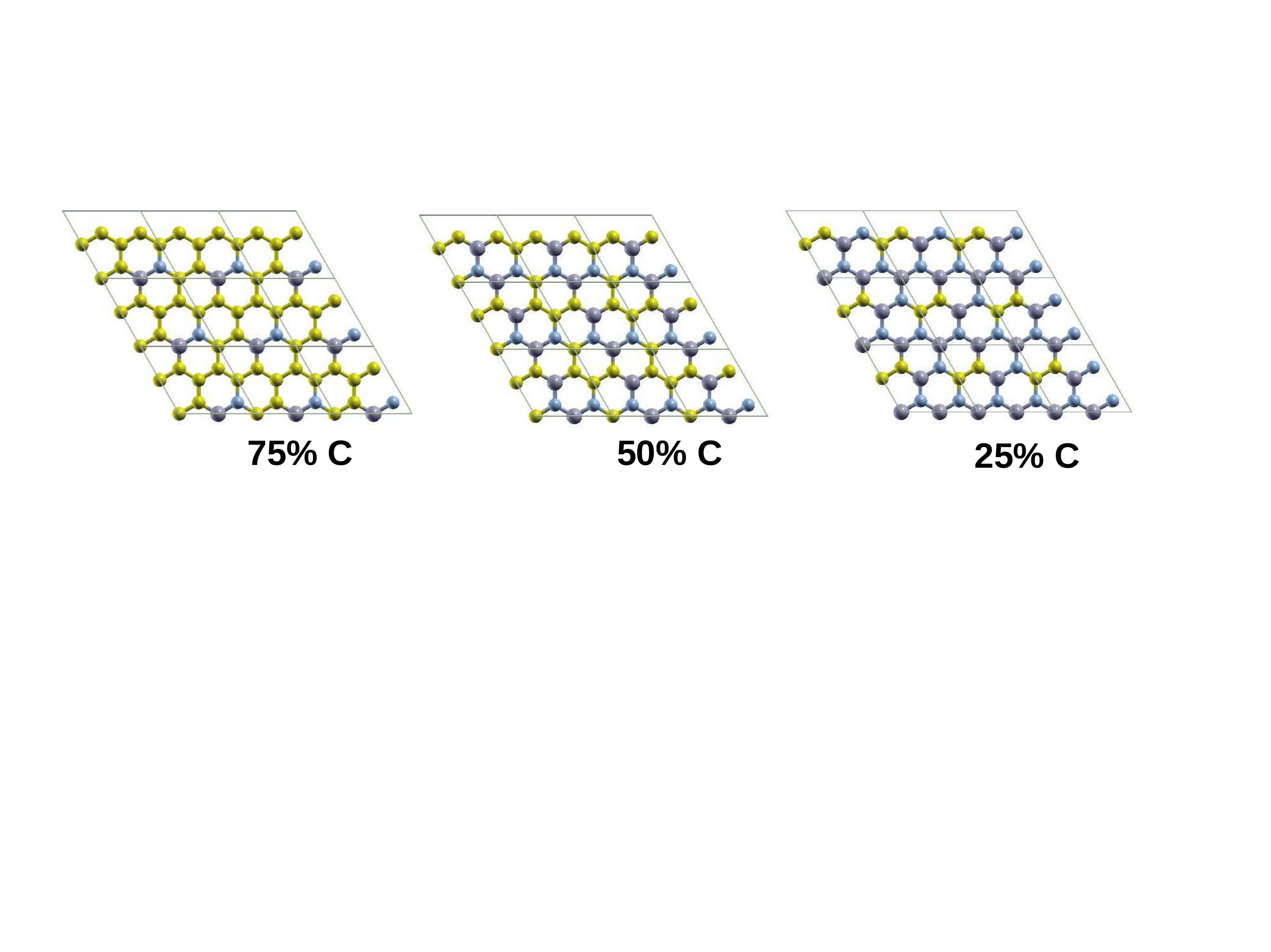}
\caption{A $2\times2$ unit cell of graphene codoped with equal number of
B and N atoms with carbon concentrations of  75\%~ (left), 50\%~ (middle)
and 25\%~ (right) used in the calculation.}
\label{fig:1}     
\end{figure}

The calculations were performed using Quantum Espresso code \cite{Ref13}. The code has already been used successfully for graphene 
and $h$-BN systyems \cite{udo1,udo2,udo3,kaloni1}. The atomic positions and cell parameters are 
fully relaxed in all cases, until an energy convergence of 10$^{-7}$ eV and a force convergence of 0.04 eV/\AA\ is reached. 
First, we obtained the total energy of pristine graphene using self-consistent calculation and then the band structure, densities of states (DOS), 
partial density of states (PDOS) and charge densities \cite{Ref15}. We have checked different sets of pseudopotentials and find von Barth \cite{Ref14} pseudopotentials is 
reasonable good for our present calculations.  Hence, we used von Barth with wave function and charge-density 
cut-offs of 60~Ryd and 600~Ryd, respectively and obtained quite accurate vales of in-plane and out-of-plane lattice constant $a$ and $c$, 
respectively by the process of the total energy minimization. 

After getting the relaxed structures, we performed self consistent calculations with a Monkhorst-Pack \cite{monk} $8\times8\times8$ k-mesh 
followed by the non-self consistent calculations for band structures, DOS/PDOS and charge density separately. We have used 
$61\times61\times61$ k-points mesh along the path $\Gamma-K-M-\Gamma$ in the irreducible Brillouin zone to obtain the band structure 
with very fine mesh points. Next, 
the doping of graphene was carried out using a $2\times2$ supercell of graphene with one atom replaced by B or N and the band structure 
was obtained without changing the lattice constant. However, ionic relaxation has been taken into account and finally, we performed 
band structures and DOS/PDOS calculations. Turning in to the multilayered structure, we first considered B or N 
doped two-layer (ABAB stacked) graphene with different concentrations of B and N atoms. Thereafter, we performed band structures and 
DOS/PDOS calculations as similar as discussed above. Finally, we have performed the relaxation, band structures and density 
of states calculations for three-layers of B or N doped graphene with similar stacking as above by keeping the interlayer spacing same as in graphite 
calculated earlier \cite{Ref15}. Based on L\"owdin population analysis, it is confirmed that in B-doped case the charge is 
concentrated on the C sites whereas, in N-doped case the charge is concentrated in N site.

Thereafter, we carried out calculations for co-doping of B and N using a $2\times2$ supercell of graphene by replacing C atoms by one, 
two or three B and same number of N atoms, yielding a concentration of 75\%, 50\% or 25\% of C atoms and corresponding concentration of 25\%,
50\% and 75\% of BN atoms, respectively.
The lattice was allowed to relax and the
in-plane lattice constant was obtained self-consistently by minimizing the total energy 
for each of these configurations. The cut-off energies in this case for wave function and charge-density were 100~Ryd and 1000~Ryd
respectively. The band structure was obtained for ($61\times61\times61$) $k$-points along the path $\Gamma-K-M-\Gamma$ exactly similar 
process as mentioned above. Similar kinds of study was also performed for multilayers (ABAB-stacked) of such BNC sheet with interlayer 
separation 3.29~\AA, which is average of the interlayer separations of pristine graphite and pristine $h$-BN calculated using same 
pseudopotentials as discussed above \cite{Ref10}.

\section{Results and Discussion}
\begin{figure*}
\includegraphics[width=0.8\textwidth]{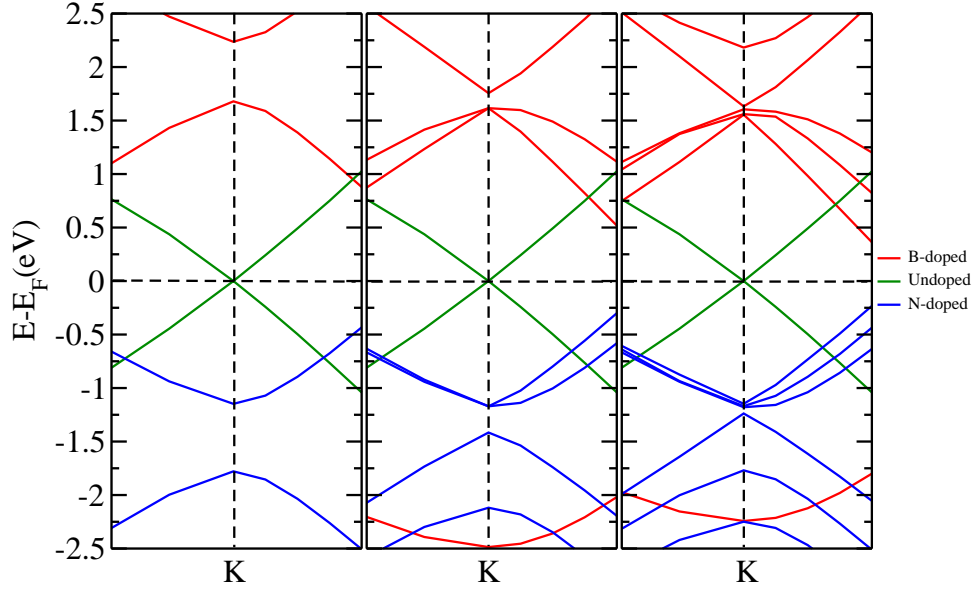}
\caption{Calculated band structure of undoped (green), B-doped (red) and N-doped (blue) graphene
calculated using LSDA, for single layer (left panel), two-layers (middle panel) and for three-layers
(right panel), shown near the high-symmetric K-point of hexagonal Brillouin zone.}
\label{fig:2}      
\end{figure*}

\begin{figure*}
\includegraphics[width=1.0\textwidth]{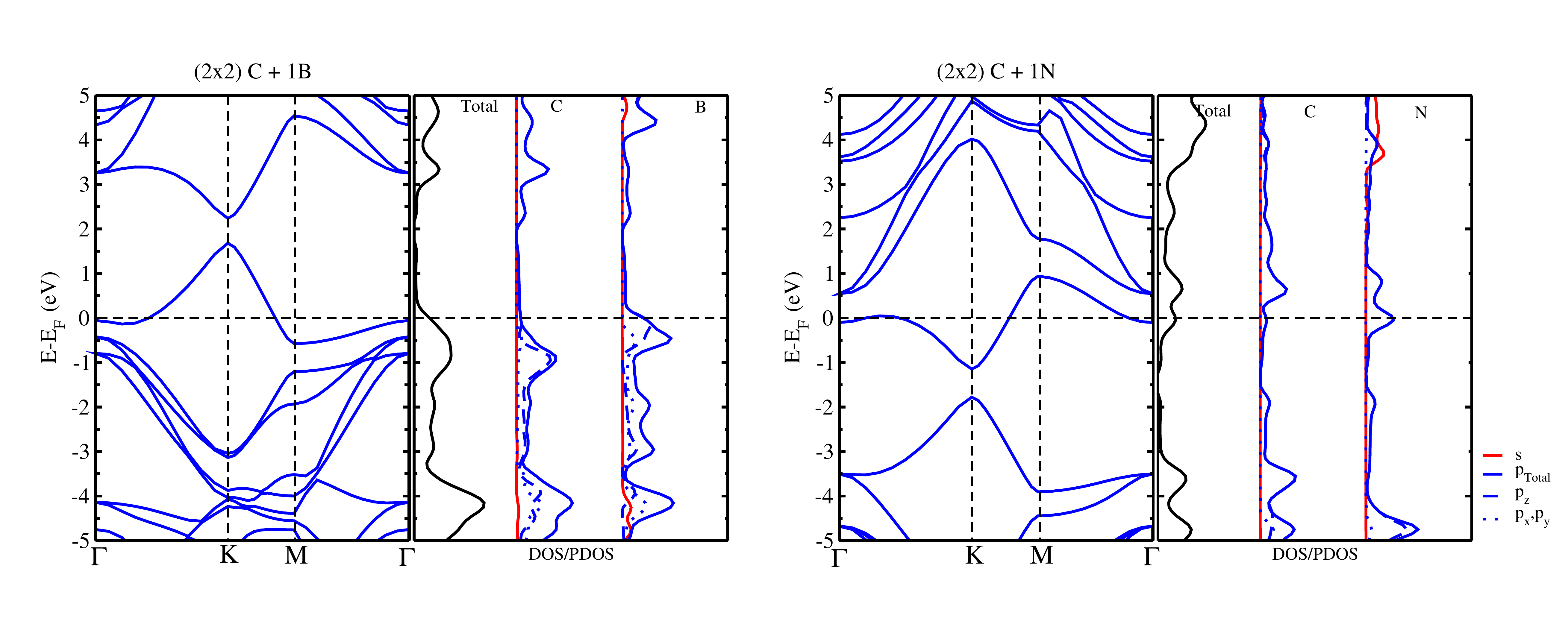}
\caption{Calculated band structure, DOS and PDOS for Boron-doped (left panel) and for Nitrogen-doped graphene (right panel).}
\label{fig:2}     
\end{figure*}


In the Fig. 2, we show the calculated band structure of undoped, B-doped and N-doped graphene for a single-layer, 
two-layers and three-layers, respectively. We clearly observe that the Dirac point for undoped graphene moves above (below) 
the  $E_F$ upon doping by B (N). B ($2s^22p^1$) has one valence electron less whereas 
N ($2s^22p^3$) has one more than Carbon ($2s^22p^2$). Therefore, doping by B (N) results as 
hole (electron) doping of graphene resulting into shifting of the bands to accommodate extra hole (electron). 
For B doped graphene, we show the band structure together with the DOS(PDOS) in left panel of the Fig.\ 3. The Dirac cone shifts above at  
$E_F$ by 2 eV with finite splitting and the linear dispersion of pristine graphene disappears. Above the $E_F$ the $\pi$ band 
contributed from the $p_z$ orbital of the B and $\pi^*$ band contributed from the $p_z$ orbital of C. However, for N doped 
graphene we obtain the band structure together with the DOS(PDOS) as shown in right panel of the Fig.\ 3. The Dirac cone  shifts below the $E_F$ 
by 1.5 eV with finite splitting. Below the E$_F$ the $\pi$ band contributed from the $p_z$ orbital of the C and $\pi^*$ band 
contributed from the $p_z$ orbital of N. In the electron doping more states are pulled below the $E_F$ whereas for hole doping 
the reverse is the case.

\begin{figure*}
\includegraphics[width=0.8\textwidth]{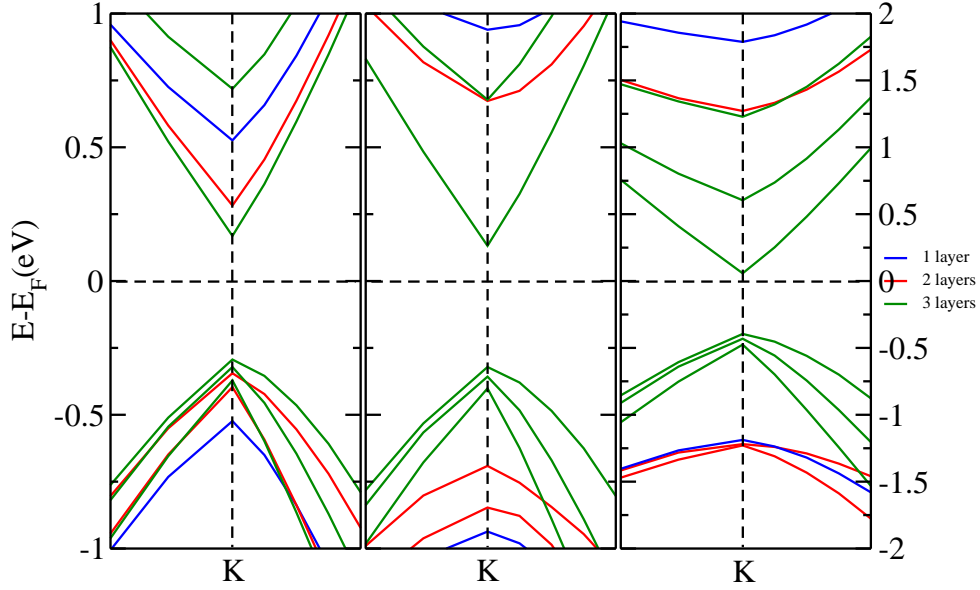}
\caption{Calculated band structure of graphene co-doped with B and N atoms at three different
concentration of C atoms: 75\% (left panel), 50\% (middle panel) and 25\% (right panel), respectively.
Results for single layer (blue), two layers (red) and three layers (green) are given.}
\label{fig:2}      
\end{figure*}

In the Fig.\ 4, we show the band structure of graphene co-doped with both B and N. The number of B and N atoms 
were kept same and calculations were performed for three configurations with concentration of C atoms to be 
75\%, 50\% and 25\%, respectively. We clearly observe the appearance of a band gap (summarized in Table I) at the high-symmetric K-point
in the Brillouin zone, 
which depends sensitively on the doping-concentration  and also on the thickness of the layers. We find for each C concentration, 
the band gap decreases with increasing number of layers. Also, with increasing B and N atom concentration the band 
gap increases. The calculated band gaps for 75\% C concentration are 1.06 eV, 0.6 eV and 0.46 eV; for 50\% C concentration 
are 1.8 eV, 1.36 eV and 0.45 eV; and for 25\% C concentration are 3.05 eV, 2.49 eV and 0.44 eV; for one-layer, two-layers 
and three-layers samples respectively. The gap occurs due to mixed hybridization of valence states of B and N with 
that of C atoms. These results are found to be in qualitative agreement with experimental measurement. The calculated value 
of band gap presented in Table I shows that such  nano-materials may have potential application in nano-scale semiconducting 
and nano-scale optoelectronic devices.

\begin{table*}[p]
\begin{tabular}{|c|c|c|c|}
\hline
Concentration (\%) & Gap$_1$ (eV) & Gap$_2$ (eV) & Gap$_3$ (eV) \tabularnewline    
\hline
\hline
$75$               & 1.06         &0.60          & 0.46 \tabularnewline
\hline
$50$               &1.80          &1.36          & 0.45 \tabularnewline
\hline
$25$               &3.05          &2.49          & 0.44 \tabularnewline
\hline
\end{tabular}\caption{Concentration of C atoms and the calculated band gap in $h$-CBN : one layer (Gap$_1$), two layers (Gap$_2$) and three layers (Gap$_3$).}
\end{table*}

We have examined the origin of the gap in the band structure due to B or N doping by calculating the
PDOS on different atomic sites, which is not shown here. We have found that the
DOS near the gap is essentially of $p_z$ character occurring from the anti-bonding (bonding) $p_z$ B (N) 
states hybridizing with those of bonding (anti-bonding) $p_z$ states of Carbon. The position of the energy gap
due to such hybridization is very crucial such that upon co-doping of graphene by equal number of B and
N the energy gap occurs at the $E_F$. Thus a band gap may be engineered in graphene by doping in 
equal amount of electron and hole. 

\section{Conclusion}

In summary, we used $ab$-initio density functional theory to investigate the effect of the B and N doped graphene. 
Shifting and splitting of the Dirac cone below and above the E$_F$ by N and B doping is observed. The band structures, 
DOS/PDOS and charge density for $h$-BCN hybrid structures are explicitly studied. We found that the band gap energy increases 
quadratically with respect to the BN concentrations in the range 25\% to 75\% keeping B to N ratio same. These results 
may provide guidance in practical engineering applications, specially to tune the band gap in graphene. Finally, our results are 
useful to provide an explain of the formation of $h$-BNC nano-materials \cite{Ref6}. These new form of hybridized $h$-BNC 
material enables for the development of band gap engineering and applications, in particular, in nano-electronics and nano-optics. 
A further study should be performed in order to shed light on this issue, in particular energy path and activation barrier.

\end{document}